\documentclass[aps,floatfix,twocolumn]{revtex4}
\usepackage{graphics,epsfig}
\usepackage{graphicx}
\begin{document}
\title{Static Ricci-flat 5-manifolds admitting the 2-sphere }
\author{Kayll Lake \cite{email}}
\affiliation{Department of Physics, Queen's University, Kingston,
Ontario, Canada, K7L 3N6 \\}
\date{\today  }
\begin{abstract}
We examine, in a purely geometrical way, static Ricci-flat
5-manifolds admitting the 2-sphere and an additional
hypersurface-orthogonal Killing vector. These are widely studied
in the literature, from different physical approaches, and known
variously as the Kramer - Gross - Perry - Davidson - Owen
solutions. The 2-fold infinity of cases that result are studied by
way of new coordinates (which are in most cases global) and the
cases likely to be of interest in any physical approach are
distinguished on the basis of the nakedness and geometrical mass
of their associated singularities. It is argued that the entire
class of solutions has to be considered unstable about the
exceptional solutions: the black string and soliton cases. Any
physical theory which admits the non-exceptional solutions as the
external vacuua of a collapsing object has to accept the
possibility of collapse to zero volume leaving behind the weakest
possible, albeit naked, geometrical singularities at the origin.
Finally, it is pointed out that these types of solutions
generalize, in a straightforward way, to higher dimensions.
\end{abstract}
\maketitle
\section{Introduction}
Wide importance is now attached to the study of higher dimensional
spaces as the arena of physical phenomena. However, as we go
beyond spacetime, the topological possibilities introduce a new
dimension to the difficulty of attaching physical significance to
various manifolds. Even at five dimensions, asymptotically flat
stationary vacuum black holes are not unique in the sense that
horizons of topology $S^1 \times S^2$ \cite{er} in addition to
$S^3$ \cite{mp} are now known. However, in the simpler static
case, it is known that asymptotically flat static vacuum black
holes are unique \cite{flat} and given by the Tangherlini
generalization of the Schwarzschild vacuum \cite{tangherlini}.
These uniqueness properties suggest that these solutions represent
the natural generalization of the Schwarzschild spacetime.
However, these are not the only asymptotically flat quasi-static
vacua which can be considered ``spherically symmetric".

Here we are concerned with 5-manifolds which admit only the 2-
sphere (unlike the Tangherlini vacua which admit the 3-sphere)
and, in addition, two hypersurface orthogonal Killing vectors at
least one of which is assumed timelike. These spaces are widely
studied in the literature from various physical approaches
\cite{wesson} but the approach used here is purely geometrical and
the results therefore applicable to any physical approach. The
only assumption made, aside from the symmetries stated, is
\begin{equation}
^{(5)}R_{a b}=0 \label{ricci}
\end{equation}
where $^{(5)}R_{a b}$ is the 5-dimensional Ricci tensor.
\section{Symmetries}
To set the notation let $\xi$ represent a Killing vector field,
\begin{equation}
\nabla_{(a} \; \xi_{\; b)}=0 \label{killing}
\end{equation}
and if, for the coordinate $x^{i}$ is adapted(
$\xi^{a}=\delta^{a}_{x^{i}}$), represent the associated field by
$_{x^{i}} \xi$. If, in addition to (\ref{killing}), $\xi$ also
satisfies the hypersurface orthogonality condition
\begin{equation}
\xi_{\; [a} \nabla_{\; b} \; \xi_{\;c]}=0, \label{hyper}
\end{equation}
write the field as $\tilde{\xi}$. The traditional definition of a
static region of spacetime is one which  admits a timelike
$\tilde{\xi}$. This terminology is also used here and we
distinguish the adapted coordinate $t$ via timelike
$_{t}\tilde{\xi}$. The spaces considered here admit an additional
hypersurface orthogonal Killing vector and we distinguish the
adapted coordinate $w$ via $_{w}\tilde{\xi}$. (Any internal
properties of $w$ are therefore of no concern.) In addition to
these symmetries we assume 2-dimensional spherical symmetry. That
is, the spaces can be decomposed as
\begin{equation}
ds^2=ds^2_3+\mathcal{R}^2d\Omega^2_2 \label{decomposition}
\end{equation}
where
\begin{equation}
d\Omega^2_2 \equiv d \theta^2+\sin(\theta)^2 d \phi^2
\label{twosphere}
\end{equation}
so that we also have the Killing field
\begin{equation}
\xi^{\theta} =c_{1} \cos(\phi)+c_{2} \sin(\phi)
\end{equation}
and
\begin{equation}
\xi^{\phi} =c_{3}-\cot(\theta)(c_{1} \sin(\phi)-c_{2} \cos(\phi))
\end{equation}
where $c_{1}, \; c_{2}$ and $c_{3}$ are non-zero constants and the
other components of this $\xi$ are zero. This $\xi$ of course
includes $_{\phi}\tilde{\xi}$. Here  $\mathcal{R}$ is independent
of $t, w, \theta$ and $\phi$ and if $\mathcal{R}=0$ we refer to
this circumstance as an ``origin".
\section{Parameter Space of Solutions}
With the foregoing symmetries it follows that (\ref{ricci}) is
satisfied by
\begin{equation}
ds^2=Ah^{\alpha}dw^2+Bh^{\delta}dt^2+\frac{dr^2+S^2
d\Omega^2_2}{fS^4} \label{sform}
\end{equation}
where $A$ and $B$ are non-zero constants, $\alpha$ and $\delta$
are constants (not both zero), $h=h(r)$ (assumed monotone),
\begin{equation}
f=C h^{\alpha+\delta-2}(h')^2 \label{f(b)}
\end{equation}
where $C$ is a non-zero constant, $^{'} \equiv d/dr$ and
\begin{equation}
S(r)=\pm\frac{h^{(2\sqrt{\epsilon \nu}-1)/(2 \sqrt{ \epsilon
\nu})}(\epsilon h^{1/\sqrt{\epsilon \nu}}-\nu)}{h^{'}}\label{S(b)}
\end{equation}
where $\epsilon$ and $\nu$ are constants restricted by the
relation
\begin{equation}
\alpha^{2}+\delta^{2}+\alpha \delta=\frac{1}{\epsilon \nu}
\equiv\frac{1}{\lambda^2}> 0. \label{consistency}
\end{equation}
To view (\ref{sform}) as static we take $B=-1$ (unless otherwise
noted) and to view (\ref{sform}) as an augmentation of spacetime
(with signature +2) we take $C>0$. We retain both signs of $A$ and
so allow the doubly static cases $A=B=-1$. Note that the forms of
$f$ and $S$ remain unchanged under the interchange $\alpha
\leftrightarrow \delta$ \cite{leon}. The reason for this symmetry
here is that the adapted coordinates $w$ and $t$ are (prior to
setting $B=-1$) interchangeable.

Since the magnitude of $A$ (and of course $B$) can be absorbed
into the scale of $w$ (and $t$), the form (\ref{sform}) would
appear at first sight to admit three independently specified
constants, in addition to $C$, via (\ref{consistency}). This,
however, is not the case as we can, without loss in generality,
set
\begin{equation}
\epsilon=\nu=\frac{1}{2} \label{onehalf}
\end{equation}
thus simplifying (\ref{S(b)}) to the form
\begin{equation}
S(r)=\pm\frac{h^2-1}{2h^{'}}.\label{Ss(b)}
\end{equation}
This is shown in the Appendix. There are then only two specifiable
parameters ($C$ and $\alpha$ or $\delta$ subject to
(\ref{consistency}) with (\ref{onehalf})).
\section{Generating explicit solutions}
Clearly the form (\ref{sform}) allows an infinite number of
representations given a monotone function $h(r)$ subject to the
reality of the resultant metric coefficients. Another approach is
to assume a form for $S(r)$ and solve the differential equation
(\ref{Ss(b)}). For example, if we set $S(r)=r$ we obtain the
solution given by  Gross and Perry \cite{grossandperry} and
Davidson and Owen \cite{davidson}. Similarly, setting
$S(r)=\sinh(r)$ we obtain the solution given recently by Millward
\cite{millward}. These representations are discussed in a more
direct way below.

Whereas (\ref{Ss(b)}) can be solved exactly for a wide variety of
choices for $S(r)$,  we can of course use $h$ as a coordinate and
view (\ref{Ss(b)}) as the required coordinate transformation. With
$h$ as a coordinate (\ref{sform}) takes the simple form
\begin{equation}
ds^2=Ah^{\alpha}dw^2+Bh^{\delta}dt^2+\frac{dh^2+\mathcal{S}^2
d\Omega^2_2 }{\mathsf{f} \mathcal{S}^4} \label{hform}
\end{equation}
where now
\begin{equation}
\mathsf{f}=\mathsf{f}(h)=(2C)^2h^{\alpha+\delta-2}, \label{f(h)}
\end{equation}
\begin{equation}
\mathcal{S}=\mathcal{S}(h)=\pm\frac{(h^2-1)}{2},\label{S(h)}
\end{equation}
and
\begin{equation}
\alpha^2+\delta^2+\alpha \delta=4. \label{condition}
\end{equation}
The coefficient in (\ref{f(h)}) has been redefined for
convenience. An important aspect of what follows is the
distinction, on geometrical grounds, of various members of the
solution locus (\ref{condition}).
\subsection*{Other representations}
Whereas the from (\ref{hform}) is convenient (and as is shown
below, complete in most cases), coordinate transformations bring
the spaces into more familiar forms. For example, with the
transformation
\begin{equation}
h = \sqrt{1-\frac{1}{Cr}}\label{kramerform}
\end{equation}
we obtain the form given by Kramer \cite{kramer}. Similarly, with
the transformation
\begin{equation}
h = \left(\frac{4Cr+1}{4Cr-1}\right)\label{grossform}
\end{equation}
we obtain form given by given by  Gross and Perry
\cite{grossandperry} (the notation (here to there) is related by
$\delta \rightarrow-2/\alpha, \alpha\rightarrow-2\beta/\alpha$,
and $C\rightarrow1/4m$) and by Davidson and Owen \cite{davidson}
(the notation (here to there) is related by $\delta
\rightarrow-2\epsilon k, \alpha\rightarrow 2 \epsilon$, and
$C\rightarrow a/4$). The solution given recently by Millward
\cite{millward} is a very special case. It is given by
\begin{equation}
h = e^{\sqrt{3}b}\label{millward}
\end{equation}
with $\alpha=\delta=\frac{2}{\sqrt{3}}$ and $C^2=\frac{1}{3M^2}$.
\section{Properties of the solutions (\ref{hform})}
\begin{figure}[ht]
\epsfig{file=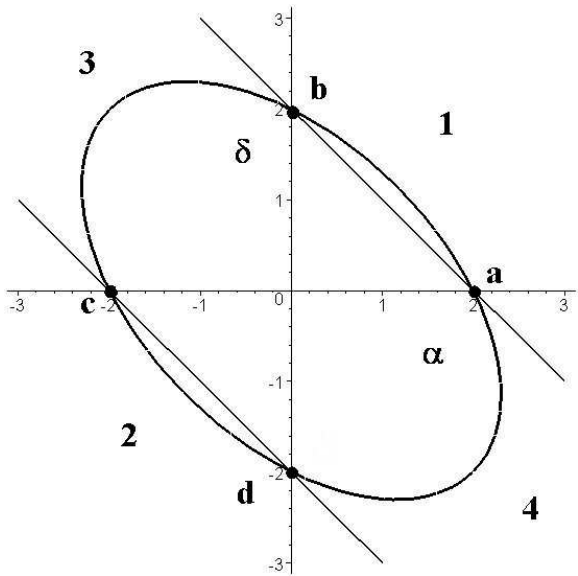,height=3in,width=3in,angle=0}
\caption{\label{locus}The solution locus $\alpha^2+\delta^2+\alpha
\delta=4$ (ellipse) along with $\alpha+\delta=2$ (top line) and
$\alpha+\delta=-2$ (bottom line). Quadrants are indicated for
convenience. These quadrants do not include the exceptional
solutions $a=(2,0), b=(0,2), c=(-2,0)$ and $d=(0,-2)$ which are
discussed separately as they have very distinct properties.}
\end{figure}

In this section we study the geometric properties of the solutions
(\ref{hform}). The various solutions are distinguished in Figure
\ref{locus} via the parameters $\alpha$ and $\delta$. First,
however, we discuss the interchange symmetry associated with the
solution locus (\ref{condition}).

\subsection*{Interchange Symmetry}
The solution locus (\ref{condition}) is obviously invariant to the
interchange $(\alpha,\delta) \leftrightarrow (-\alpha,-\delta)$
\cite{wesson1}. To see why, consider the coordinate transformation
\begin{equation}
h=\frac{1}{r}.\label{inversion}
\end{equation}
Under the transformation (\ref{inversion}) we obtain (\ref{hform})
but with $h \rightarrow r$, $\alpha \rightarrow -\alpha$, and
$\delta \rightarrow -\delta$. Moreover, in terms of Figure
\ref{locus}, we have the quadrant interchanges $1 \leftrightarrow
2$ and $ 3 \leftrightarrow 4,$ along with the point interchanges
$a \leftrightarrow c$ and $ b \leftrightarrow d$. As a result, the
form (\ref{hform}) duplicates all distinct classes of solutions.
As a matter of convenience we continue with this form here but
with knowledge of this duplicity.
\subsection*{Distinguishing The Constant $C$}
Before the properties of the solutions (\ref{hform}) are discussed
in terms of the parameters $\alpha$ and $\delta$, it is useful to
review the general nature of the constant $C$ as it is quite
distinct from $\alpha$ and $\delta$. This is similar to (the
reciprocal of) the ``mass" in the four-dimensional Schwarzschild
vacuum as one might well guess from the other representations
discussed above and as we explain in detail below. To see this let
$ds^2$ represent the 4 dimensional Schwarzschild vacuum (of mass
$m$) and subject this to a conformal transformation
\begin{equation}
ds^2 \rightarrow \Phi^2 ds^2 \label{conformal}
\end{equation}
where $\Phi$ is a constant. This, of course, preserves Ricci
flatness (in any dimension). It is easy to show that under the
transformation (\ref{conformal})
\begin{equation}
m\rightarrow\pm|\Phi|m. \label{conformalmass}
\end{equation}
It is clear from the form (\ref{hform}) then that $1/C$ plays a
role similar to $m$ (using the freedom in the scale of $w$ and
$t$). The parameters $\alpha$ and $\delta$, however, distinguish
the solutions (\ref{hform}) in a rather different way as we now
examine.
\subsection*{Origins}
As explained above, we refer to $\mathcal{R}^2=0$ as an origin.
From (\ref{hform}) we have
\begin{equation}
\mathcal{R}^2=\frac{h^{2-\alpha-\delta}}{C^2(h^2-1)^2}\label{origin}
\end{equation}
so that in all cases
$\lim_{h\rightarrow1^{\pm}}\mathcal{R}^2\rightarrow\infty$.
Moreover, $ \mathcal{R}$ has a minimum ($>0$) at
\begin{equation}
h^2=\frac{\alpha+\delta-2}{\alpha+\delta+2}\label{Rmin}
\end{equation}
which restricts the minima to quadrants $1$ and $2$ of Figure
\ref{locus}. In quadrant $1$ the minima occur for $h\in(0,1)$ and
in quadrant $2$ they exist for $h\in(1,\infty)$. For the
exceptional solutions $a$ and $b$ clearly
$\mathcal{R}^2=1/C^2(h^2-1)^2$ and for the exceptional solutions
$c$ and $d$, $\mathcal{R}^2=h^4/C^2(h^2-1)^2$. Properties of
$\mathcal{R}^2$ are summarized in Tables \ref{tab:regular} and
\ref{tab:exceptional}.
\begin{table}
\caption{\label{tab:regular}``Regular" solutions}
\begin{tabular}{|c|c|c|c|r|}
    \hline
 $h $&  1    &   2  & 3 and 4\\
    \hline
$\in(0,1)$   & $\mathcal{R}^2\rightarrow\infty$& $\mathcal{R}^2\rightarrow0$ & $\mathcal{R}^2\rightarrow0$  \\
 $\rightarrow0^{+}$& $W\rightarrow\infty$ & $W\rightarrow\infty$ &  $W\rightarrow\infty$    \\
  $\;$ & $SL^{*}$ & $TL$ & $N$ and $TL$    \\
  $\;$& $m\rightarrow \pm \infty$ & $m \rightarrow 0$    &   $m \rightarrow \pm \infty$\\
  \hline
$\rightarrow1^{\pm}$   & $\mathcal{R}^2\rightarrow\infty$& $\mathcal{R}^2\rightarrow\infty$& $\mathcal{R}^2\rightarrow\infty$\\
& $W\rightarrow0$&$W\rightarrow0$&$W\rightarrow0$\\
  \hline
$\in(1,\infty)$   & $\mathcal{R}^2\rightarrow0$ &$\mathcal{R}^2\rightarrow\infty$&$\mathcal{R}^2\rightarrow0$\\
$\rightarrow\infty$& $W\rightarrow\infty$ &$W\rightarrow\infty$&$W\rightarrow\infty$\\
  $\;$ & $TL$ & $SL^{*}$  & $TL$ and $N$    \\
   $\;$& $m\rightarrow 0$ & $m \rightarrow \pm \infty$    &   $m \rightarrow \pm \infty$\\
    \hline
\end{tabular}
\end{table}

\begin{table}
\caption{\label{tab:exceptional}``Exceptional" solutions}
\begin{tabular}{|c|c|c|r|}
    \hline

$h $&  $a$ and $b$    &   $c$ and $d$ \\
    \hline
$\in(0,1)$& $\mathcal{R}^2\rightarrow1/C^2$& $\mathcal{R}^2\rightarrow0$\\
$\rightarrow0^{+}$   & $W\rightarrow12C^4$ & $W\rightarrow\infty$ \\
$\;$   & $\;$ & $TL$ and $TL$ \\
 \hline
$\rightarrow1^{\pm}$   & $\mathcal{R}^2\rightarrow\infty$& $\mathcal{R}^2\rightarrow\infty$\\
& $W\rightarrow0$&$W\rightarrow0$\\
 \hline
$\in(1,\infty)$& $\mathcal{R}^2\rightarrow0$ & $\mathcal{R}^2\rightarrow1/C^2$\\
$\rightarrow\infty$   & $W\rightarrow\infty$ &$W\rightarrow12C^4$\\
$\;$   & $TL$ and $TL$  & $\;$ \\
    \hline
\end{tabular}
\end{table}

\subsection*{Weyl Invariant}
For the spaces (\ref{hform}) there is only one independent
invariant derivable from the Riemann tensor without
differentiation and this can be taken to be $^{(5)}C_{abcd}$$
^{(5)}C^{abcd} \equiv W$ where $^{(5)}C_{abcd}$ is the
5-dimensional Weyl tensor. This is given by \cite{weyl}
\begin{equation}
W=\frac{3h^{2(\alpha+\delta-4)}(h^2-1)^6C^4\mathcal{H}}{32}\label{weylsq}
\end{equation}
where
\begin{eqnarray}
\mathcal{H}\equiv(8-\alpha \delta)\left((8+\alpha
\delta)(h^4+1)+4(\alpha+\delta)(h^4-1)\right)\nonumber\\ +2h^2
\alpha^2 \delta^2.  \;\;\;\;\;\                  \label{Hform}
\end{eqnarray}

First note that for the exceptional cases $a$ and $b$
\begin{equation}
W=12(h^2-1)^6C^4, \label{weylsqex1}
\end{equation}
and for the exceptional cases $c$ and $d$
\begin{equation}
W=\frac{12(h^2-1)^6C^4}{h^{12}}. \label{weylsqex2}
\end{equation}
In all cases
\begin{equation}
\lim_{h\rightarrow 1^{\pm}}W\rightarrow0.\label{hone}
\end{equation}
Moreover,
\begin{equation}
\lim_{h\rightarrow0}W\rightarrow\infty \label{weylsingzero}
\end{equation}
except for the exceptional cases $a$ and $b$ for which
$\lim_{h\rightarrow0}W\rightarrow 12C^4$  and
\begin{equation}
\lim_{h\rightarrow \infty}W\rightarrow\infty \label{weylsing}
\end{equation}
except for the exceptional cases $c$ and $d$ for which
$\lim_{h\rightarrow  \infty}W\rightarrow 12C^4$. The limit
(\ref{weylsingzero}) shows that in all but the two cases
indicated, the solution (\ref{hform}) is singular at $h=0$.
Similarly, the limit (\ref{weylsing}) shows that in all but the
other two cases indicated, the solution (\ref{hform}) is singular
as $h\rightarrow  \infty$. The properties discussed above are
summarized in Tables \ref{tab:regular} and \ref{tab:exceptional}.
\subsection*{Local Asymptotic flatness}
Since, as shown above,
\begin{equation}
\lim_{h\rightarrow1^{\pm}}\mathcal{R}^2\rightarrow\infty,
\label{asymp}
\end{equation}
and since
\begin{equation}
\lim_{h \rightarrow 1^{\pm}}\; ^{(5)}R_{ab}^{\;\;\; cd}
\rightarrow 0, \label{Riemann}
\end{equation}
where $^{(5)}R_{ab}^{\;\;\; cd}$ is the 5-dimensional Riemann
tensor, all  solutions are (locally) asymptotically flat for $h
\rightarrow 1^{\pm}$.
\subsection*{Subspaces $w$ $=$ constant}
In any subspace $w=$ constant it follows from (\ref{hform}) that
$^{(4)}R_{a b}=0$ only for $\alpha=0$, that is a subspace of the
exceptional solutions $b$ and $d$. These are ``black string"
solutions \cite{string}. Moreover, $C^2=1/4m^2$ where $m$ is the
usual Schwarzschild mass (as we discuss further below). With the
aide of Table \ref{tab:exceptional} we see that for the solution
$b$, $h\in(0,1)$ covers the static part of the spacetime with
$m>0$ and all of the spacetime with $m<0$ for $h\in(1,\infty)$.
Similarly, the solution $d$ with $h\in(1,\infty)$ covers the
static part of the spacetime with $m>0$ and all of the spacetime
with $m<0$ for $h\in(0,1)$. (In an analogous way, we can consider
any subspace $t=$ constant with $A=-1$ and interchange $a$ for
$b$, $c$ for $d$ and $\alpha$ for $\delta$ in the foregoing. The
exceptional cases $a$ and $c$ we call ``soliton" solutions
\cite{grossandperry}.)
\subsection*{Nature of the Singularities}
Hypersurfaces of constant $h$ in subspaces of constant $w$ can be
categorized by way of the trajectories they contain. Here we
distinguish the spacelike ($SL, 2\mathcal{L}>0$), timelike ($TL,
2\mathcal{L}<0$) and null ($N, 2\mathcal{L}=0$) cases where $2
\mathcal{L} \equiv \;^{(5)}v_{a}\;^{(5)}v^{a}$ and
$^{(5)}v^{a}=(0,\dot{t},0,\dot{\theta},\dot{\phi)}$ is tangent to
a curve in the hypersurface ($^{.} \equiv d/d\lambda$, where
$\lambda$ is any parameter that distinguishes events along the
curve). From (\ref{hform}) we find
\begin{equation}
2
\mathcal{L}=-h^{\delta}\dot{t}^2+\mathcal{R}^2(\dot{\theta}^2+\sin(\theta)^2
\dot{\phi}^2). \label{2l}
\end{equation}
The nature of the singularities thus categorized are summarized in
Tables \ref{tab:regular} and \ref{tab:exceptional} where $SL^{*}$
means $N$ if exactly radial ($\dot{\theta}=\dot{\phi}=0$).
\subsection*{Null Geodesics}
Write the momenta conjugate to $_{t}\tilde{\xi},
\;_{w}\tilde{\xi}$ and $_{\phi}\tilde{\xi}$ as $p_{t},\; p_{w}$
and $p_{\phi}$ and define the constants of motion $\mathcal{C}_{w}
\equiv p_{w}/p_{t}$ and $\mathcal{C}_{\phi} \equiv
p_{\phi}/p_{t}$. The null geodesic equation now follows as
\begin{equation}
\dot{h}^2=f\mathcal{S}^4\left(\frac{1}{h^{\delta}}-\frac{\mathcal{C}_{w}^2}{A
h^{\alpha}}-\frac{\mathcal{C}_{\phi}^2}{\mathcal{R}^2}\right)
\label{null}
\end{equation}
where $^{.} \equiv d/d \lambda$ for a suitably scaled affine
parameter $\lambda$ and without loss in generality we have set
$\theta=\pi/2$. In general then the singularities discussed above
are ``directional" in their visibility as there exist turning
points in $h$. In the ``radial" case
$\mathcal{C}_{w}=\mathcal{C}_{\phi}=0$ we have
\begin{equation}
\dot{h}^2=h^{\alpha-2}(h^2-1)^4. \label{radialnull}
\end{equation}
Whereas (\ref{radialnull}) can be integrated in terms of special
functions, this integration is not needed.  We simply observe that
there are no turning points on $h \in (0,1)$ or on $h \in
(1,\infty)$. As a result, only in the exceptional solutions $a$
and $b$ on $h \in (0,1)$ and $c$ and $d$ on $h \in (1,\infty)$ are
the forms (\ref{hform}) incomplete and also free of naked
singularities in the ranges given \cite{naked}.
\subsection*{Static Extensions}
It is clear from Table \ref{tab:exceptional} that the exceptional
solutions $a$ and $b$ are incomplete on $h \in (0,1)$ and the
exceptional solutions $c$ and $d$ are incomplete on $h \in
(1,\infty)$. Since there is no geometrical reason to terminate
these cases, we consider here simple static extensions. In these
cases (but clearly not in general) we can consider $h<0$. First
note that the discussion on asymptotic flatness given above
extends to $h\rightarrow-1^{\pm}$ in these cases. Moreover, the
results given in Table \ref{tab:exceptional} extend directly to
$h<0$. It follows then that the cases $a$ and $b$ on $h \in(-1,1)$
and $c$ and $d$ on $h \in(-1,-\infty)\bigcup(1,\infty)$ are
singular-free. However, as we know from the 4-dimensional case,
these static extensions can be incomplete.
\subsection*{Non-Static Extensions}
For the black string solutions $b$ and $d$ we can write
\begin{equation}
ds^2=Adw^2+d\tilde{s}^2 \label{stringextension}
\end{equation}
where $d\tilde{s}^2$ is any regular completion of the
Schwarzschild manifold so the space is no longer static via
$_{t}\tilde{\xi}$. If $A=-1$ the space remains static via
$_{w}\tilde{\xi}$. Similarly, if $A=-1$ for the soliton solutions
$a$ and $c$ we can write
\begin{equation}
ds^2=-dt^2+d\tilde{s}^2 \label{stringextension1}
\end{equation}
where again $d\tilde{s}^2$ is any regular completion of the
Schwarzschild manifold so the space is no longer static via
$_{w}\tilde{\xi}$ but remains static via $_{t}\tilde{\xi}$. Of
course, the past spacelike singularities in $d\tilde{s}^2$ render
these solutions also nakedly singular, but in a way quite unlike
the foregoing.
\subsection*{Geometrical Mass}
Not all singularities can be considered equally serious. Here we
define the geometrical mass associated with spherical symmetry in
order to classify the singularities discussed above. Define the
quantity $m$ via the sectional curvature of the two-sphere
\cite{mass},
\begin{equation}
m \equiv \frac{^{(5)}R_{\theta \phi}^{\;\;\;\;\theta
\phi}g_{\theta \theta}^{3/2}}{2}. \label{mass}
\end{equation}
(In 4 dimensions the well known effective gravitational mass of
spherically symmetric spacetimes is given by (\ref{mass}) with
$^{(5)}$ replaced by $^{(4)}$.)

It follows from (\ref{hform}) that \cite{masswesson}
\begin{equation}
m =
-\frac{((\alpha+\delta)^2+4)(h^2-1)+4(\alpha+\delta)(h^2+1)}{32Ch^{(\alpha+\delta+2)/2}},
\label{masshform}
\end{equation}
so that in all cases
\begin{equation} \lim_{h\rightarrow1}m =
-\frac{\alpha+\delta}{4C}. \label{masshone}
\end{equation}
However, only in the exceptional solutions is $m$ a constant and
then,
\begin{equation}
m=\pm\frac{1}{2C}. \label{massC}
\end{equation}
The properties of $m$ in the other solutions are summarized in
Table \ref{tab:regular}.  In what follows we classify the
geometrical ``strength" of a singularity on the basis of $m$.
\section{The View from Above}
Most of the foregoing generalizes in a straightforward way to
higher dimensions. For example, at dimension 6, introducing
another hypersurface-orthogonal Killing vector $_{y}\tilde{\xi}$,
it follows (retaining the previous notation) that
\begin{equation}
ds^2=Ah^{\alpha}dw^2+Bh^{\delta}dt^2+Dh^{\epsilon}dy^2+\frac{dh^2+\mathcal{S}^2
d\Omega^2_2 }{\mathsf{f} \mathcal{S}^4}, \label{hformsix}
\end{equation}
where $D$ and $\epsilon$ are constants,
\begin{equation}
\mathsf{f}=(2C)^2h^{\alpha+\delta+\epsilon-2}, \label{f(h)}
\end{equation}
and (\ref{S(h)}) still holds, satisfies $^{(6)}R_{a}^{b}=0$ as
long as the parameters remain on the ellipsoid
\begin{equation}
\alpha^2+\delta^2+\epsilon^2+\alpha \delta+\alpha \epsilon +
\delta \epsilon =4. \label{conditionsix}
\end{equation}
Of particular interest here is the fact that in any subspace $y =
$ constant it follows from (\ref{hformsix}) that
$^{(5)}R_{a}^{b}=0$ only for $\epsilon=0$. These subspaces are
precisely the spaces considered in this paper.

\section{Discussion}
As the foregoing discussion makes clear, solutions on the locus
(\ref{condition}) have quite distinct geometrical properties, and
not all of these solutions can be considered equally valid in any
physical approach. For example, solutions in the quadrants 3 and 4
have the strongest possible geometrical singularities at the
origin ($m$ diverges at $\mathcal{R}=0$), and these singularities
are visible throughout the associated spaces for the entire range
in $h$. In quadrant 1 for $h \in (0,1)$ and quadrant 2 for $h \in
(1, \infty)$ the solutions not only lack an origin, they have the
strongest possible geometrical singularities. In contrast,
quadrant 1 solutions with $h \in (1,\infty)$ and quadrant 2
solutions with $h \in (0,1)$ have visible but the weakest possible
geometrical singularities at the origin ($m\rightarrow0$). In
contrast, the exceptional solutions have quite distinct
properties. Solutions $a$ and $b$ with $h \in (1,\infty)$ and
solutions $c$ and $d$ with $h \in (0,1)$ have naked singularities
at the origin but of intermediate strength ($m$ remains finite
\cite{deformed}). Only for the solutions $a$ and $b$ with $h \in
(0,1)$ and the solutions $c$ and $d$ with $h \in (1,\infty)$ are
the coordinates used in (\ref{hform}) incomplete. In these cases
singularities in the non-static extensions are obtained. Because
of the very distinct properties of the solutions as one covers the
solution locus (\ref{condition}), the natural conclusion is that
the solution (\ref{hform}) is fundamentally unstable about these
exceptional solutions. That is, the properties of the exceptional
black string and soliton solutions are unstable to any metric
perturbation. In general, if one was to envisage a physical theory
in which (\ref{hform}) was the external geometry of an object
collapsing to zero volume, quadrant 1 solutions with $h \in
(1,\infty)$ (equivalently quadrant 2 solutions with $h \in (0,1)$)
would seem to be the natural choice since their properties are
stable (away from the exceptional solutions) and have the weakest
possible, albeit naked, geometrical singularities at the origin
\cite{zeromass}. Finally, it was pointed out how the spaces
considered here can be thought of as subspaces of a higher
dimensional Ricci-flat manifold of similar type.

\bigskip

\begin{acknowledgments}
This work was supported by a grant from the Natural Sciences and
Engineering Research Council of Canada and was made possible by
use of \textit{GRTensorIII} \cite{grt}.
\end{acknowledgments}

\bigskip

\appendix*
\section{$\epsilon=\nu=\frac{1}{2}$}
In this appendix we prove that we can, without loss in generality,
set $\epsilon=\nu=\frac{1}{2}$ in (\ref{sform}) thus simplifying
(\ref{S(b)}) to the form (\ref{Ss(b)}). We include $B$ for
completeness. Define
\begin{equation}
\bar{\alpha} \equiv 2 \lambda \alpha, \;\;\; \bar{\delta} \equiv 2
\lambda \delta \label{newalpha}
\end{equation}
so that
\begin{equation}
\bar{\alpha}^2 +\bar{\delta}^2
+\bar{\alpha}\bar{\delta}=4.\label{newrelation}
\end{equation}
Next, define $H$ by
\begin{equation}
h \equiv \left(\frac{\nu}{\epsilon}\right)^{\lambda}H^{2
\lambda}.\label{newh}
\end{equation}
With these definitions (\ref{sform}) transforms as follows:
\begin{equation}
Ah^{\alpha} \rightarrow \bar{A}H^{\bar{\alpha}}, \;\;\;
Bh^{\delta} \rightarrow \bar{B}H^{\bar{\delta}} \label{AB}
\end{equation}
where
\begin{equation}
\bar{A}=A\left(\frac{\nu}{\epsilon}\right)^{\bar{\alpha}/2},\;\;\;\bar{B}=B\left(\frac{\nu}{\epsilon}\right)^{\bar{\delta}/2}
\label{ABnew}
\end{equation}
and
\begin{equation}
f \rightarrow \bar{C}H^{\bar{\alpha}+\bar{\delta}-2}(H^{'})^2
\label{C}
\end{equation}
where
\begin{equation}
\bar{C}=4\lambda^2\left(\frac{\nu}{\epsilon}\right)^{(\bar{\alpha}+\bar{\delta})/2}C\label{Cnew}
\end{equation}
and finally
\begin{equation}
\mathcal{S}\rightarrow\pm\frac{H^2-1}{2H^{'}}.\label{Snew}
\end{equation}
Thus, with the relabelling $H \rightarrow h$ and the removal of
the $^{\bar{}}\;\;$, we obtain a space equivalent to the form
(\ref{sform}) but with $\epsilon=\nu=\frac{1}{2}$.


\begin{thebibliography}{}\label{sec:TeXbooks}
\bibitem[*]{email}{Electronic Address: lake@astro.queensu.ca}
\bibitem{er}R. Emparan and H. S. Reall, Phys. Rev. Lett. {\bf 88}, 101101
(2002) (hep-th/0110260).
\bibitem{mp}R. C. Myers and M. J. Perry, Annals Phys. {\bf 172}, 304 (1986) .
\bibitem{flat} S. Hwang, Geometriae Dedicata {\bf 71}, 5 (1998), G.
W. Gibbons, D. Ida and T. Shiromizu, Prog. Theor. Phys. Suppl.
\textbf{148}, 284 (2003) {\tt arXiv:gr-qc/0203004}
\bibitem{tangherlini}F. R. Tangherlini, Nuovo Cimento {\bf 27}, 636 (1963).
Explicit regular coordinates that cover all of the Tangherlini
solutions are given by K. Lake, JCAP 0310 (2003) 007 {\tt
arXiv:gr-qc/0306073}
\bibitem{wesson}See, for example, J. M. Overduin and P. S. Wesson, Phys. Rept. \textbf{283}
(1997) {\tt arXiv:gr-qc/9805018}.
\bibitem{leon} For further discussion of this property see J. Ponce de Leon, Class. Quantum Grav. \textbf{23}
3043 (2006) {\tt arXiv:gr-qc/0512067}
\bibitem{grossandperry}D. J. Gross and M. J. Perry, Nucl. Phys. \textbf{B 226,} 29
(1983). Our use of the term ``soliton" here is consistent with
this work.
\bibitem{davidson}A. Davidson and D. Owen, Phys. Lett. \textbf{B 155}, 247 (1985).
\bibitem{millward}R. S. Millward {\tt arXiv:gr-qc/0603132 }
\bibitem{kramer}D. Kramer, Acta Phys. Polon. \textbf{B2}, 807 (1970).
\bibitem{wesson1} The significance of this symmetry (along with
$\alpha\leftrightarrow\delta$ explained previously) has sometimes
been considered obscure. See, for example, P. S. Wesson,
\textit{Space-Time-Matter} (World Scientific, Singapore, 1999).
\bibitem{weyl}This invariant has been given previously in different coordinates (see, for
example, the references in \cite{wesson}), but to our knowledge
the discussion given here is more complete than any given
previously. Moreover, the evolution in $W$ must be considered
along with the geometrical mass $m$ as defined below.
\bibitem{string}It is well known that the black string solutions
are unstable to large-scale perturbations. See R. Gregory, Class.
Quantum Grav., \textbf{17}, L125 (2000) {\tt arXiv:hep-th/0004101}
\bibitem{naked}In the quadrants 1 and 2 it might be argued that
the $SL^{*}$ singularities could be entirely to the future and
therefore not visible. However, since these degenerate to (single)
null singularities in the radial direction, they are necessarily
naked.
\bibitem{mass}This is a special case of the geometrical mass in
$n$ dimensions for spaces admitting a $D$-sphere:
\begin{eqnarray}
m \equiv \frac{^{(n)}R_{\theta \phi}^{\;\;\;\;\theta
\phi}g_{\theta \theta}^{(1+D)/2}}{2}.\nonumber
\end{eqnarray}
See K. Lake {\tt arXiv:gr-qc/0507031} (to be updated).
\bibitem{masswesson} The mass defined in \cite{wesson} (and in
particular see P. S. Wesson and J. Ponce de Leon, Class. Quantum
Grav. \textbf{11}, 1341 (1994)) is not equivalent to the
geometrical mass (\ref{mass}). The mass used in \cite{wesson} is,
in our notation, $-\delta/4 C h^{\alpha/2}$.
\bibitem{deformed} The fact that $m\rightarrow0$ in quadrants 1
ans 2 but $m$ remains strictly non-zero in the exceptional
solutions means that no limiting procedure can be invoked.
\bibitem{zeromass}Zero mass naked singularities arise in standard
general relativity. See K. Lake, Phys. Rev. Lett. \textbf{68},
3129 (1992).
\bibitem{grt}This is a package which runs within Maple. It is entirely
distinct from packages distributed with Maple and must be obtained
independently. The GRTensorII software and documentation is
distributed freely on the World-Wide-Web from the address \textit{
http://grtensor.org} GRTensorIII software is in development.
\end{thebibliography}
\end{document}